# SIMILARITY OF SKELETAL STRUCTURES IN LABORATORY AND SPACE AND THE PROBABLE ROLE OF SELF-ASSEMBLING OF A FRACTAL DUST IN FUSION DEVICES[*]

A.B. Kukushkin and V.A. Rantsev-Kartinov

RRC "Kurchatov Institute",
Moscow, 123182, Russia

**ABSTRACT**

This papers briefly reviews the progress in studying the long-lived filamentary structures of a skeletal form (namely, tubules and cartwheels, and their simple combinations) in electric discharges in various fusion devices. These include fast Z-pinch, tokamak and laser produced plasmas. We also report on the results of a search for the phenomenon of skeletal structures -- formerly revealed in laboratory data from fusion devices -- at larger and much larger length scales, including the powerful electromagnetic phenomena in the Earth atmosphere and cosmic space. It is found that the similarity of, and a trend toward self-similarity in, the observed skeletal structures more or less uniformly covers the range $10^{-5}$-$10^{23}$ cm. These evidences suggest all these skeletal structures, similarly to skeletons in the particles of dust and hail, to possess a fractal condensed matter of particular topology of the fractal. The probable role of the phenomenon of self-assembling of a fractal dust in fusion devices and outside the fusion is discussed briefly.

## 1. INTRODUCTION

The present paper gives a brief review of

(i) the progress that have been achieved since the last Symposium in our studies of long-lived filamentary structures of a skeletal form (namely, tubules and cartwheels, and their simple combinations) in electric discharges in various fusion devices, and

(ii) the results of a search for the phenomenon, formerly revealed in laboratory data from fusion devices [1-4], at larger and much larger length scales, including the powerful electromagnetic phenomena in the Earth atmosphere and cosmic space.

The hypothesis [1,2] suggested the long-lived filaments to possess a microsolid skeleton which might be assembled during electric breakdown, prior to appearance of major plasma, from wildly produced carbon nanotubes (or similar nanostructures of other chemical elements). The proof-of-concept studies revealed the presence of tubular and cartwheel-like structures in

(i) the high-resolution (visible light and x-ray) images of plasma in tokamaks, Z-pinches, plasma focus, and laser-produced plasmas (in the range 100 μm - 10 cm) [1-4], including the images taken at electric breakdown stage of discharge in tokamak, plasma focus and vacuum spark [3,4], (ii) various types of dust deposits in tokamak T-10 (10 nm - 10 μm) [5].

---

[*] This is a shortened version of the paper to be published in «Current Trends in International Fusion Research» (Proc. 5-th Symposium, Washington D.C., USA, March 2003). Eds. C.D. Orth, E. Panarella, and R.F. Post. NRC Research Press, Ottawa, Canada.

Here we append previous evidences [3] for skeletal structuring with the evidences (Sec. 2) found in the data on imaging the fast Z-pinch [6] and the laser-produced plasmas [4], as well as on electron microscopy of dust deposits in tokamak [7(a,b)].

Then we briefly discuss the problems of diagnosing the filamentary structures in fusion plasmas (Sec. 3). These include (a) the observability of skeletal filamentary structures in plasmas of a fast Z-pinch [6], besides the observability in a slow Z-pinch [1-4], and (b) the missed diagnostic opportunities of streak camera imaging in tokamaks [8].

Further, we extend the evidences for skeletal structuring to larger length scales (Sec. 4). We present some results of analyzing the images of hail particles (1-10 cm), tornado ($10^3$-$10^5$ cm), and of a wide class of objects in space ($10^{11}$-$10^{23}$ cm), including the solar coronal mass ejection, supernova remnants, and some galaxies [9]. This will enable us to hypothesise for the probable presence of a fractal condensed matter of particular topology of the fractal in the entire range $10^{-5}$-$10^{23}$ cm.

In Sec. 5, we discuss the present status of the qualitative model [1,2], These include (i) recent findings of anomalous magnetism of, and low dissipation of trapped magnetic flux in, carbon nanotubes and their assemblies, (ii) conservation of topology in a growing/expanding skeleton, (iii) survivability of skeletons in an ambient plasma.

In Sec. 6, we briefly discuss probable role of the phenomenon of self-assembling of a fractal dust in fusion devices and outside the fusion science.

## 2. MORE EVIDENCES FOR SKELETAL STRUCTURING IN THE LABORATORY

### 2.1 Longevity of Skeletal Structuring in Fast Z-pinch

The longevity of skeletal structures in the slow gaseous Z-pinch E-2 (discharge tube 60 cm long, 20 cm diameter, energy store ~30 kJ; initial voltage ~ 30 kV; initial pressure of deuterium gas 1.2 Torr; maximal current ~150 kA) has been demonstrated via tracing the dynamics of the few centimeters long, straight filaments directed nearly transverse to major electric current of the Z-pinch (see Figure 2 in [1] or Fig. 1 in [2(b)]). Those images were taken in the same discharge at different time moments - during about half a microsecond -- and different observation angles θ in the plane orthogonal to Z-pinch axis.

Here we present the data from the fast Z-pinch facility S-300 [10]. These data were obtained in the experiments on studying the fine structure of fast Z-pinches for the program of plasma x-ray radiation sources. In these discharges in the S-300, the total electric currents amounts to 1-3 MA, with the current growth front of ~ 100 ns duration. The load was a tailored cylinder (8 mm long, 3-5 mm in diameter, with a neck of ~ 1 mm diameter), made of agar-agar with various heavy-element fillings. The analyzed images were taken with the help of electronic optical converters (exposure 3 ns, interframe interval 15 ns, spatial resolution 50-100 μm). In a single discharge, from 3 to 5 images were successively taken in the visible light and/or soft x-rays (SXR), with covering a part of vacuum ultraviolet (VUV) spectral range.

The original images were processed with the help of the method of multilevel dynamical contrasting (MDC) [11(a,b)] (this method may increase spatial resolution by an order of magnitude). Basically, for the images of high spatial resolution the structures under search are practically resolvable, however, without any MDC processing: it is merely suffice to magnify the original image to a proper size. The MDC processing resolves the fine structure and makes it more distinct.

Below we present the pictures which are obtained from original images using a «homogeneous» map of contrasting, i.e. a single map for the entire image (in general, the map of contrasting is a dependence $I_l$ vs. $I_0$ which prescribes that the given value $I_0$ of image's blackening is to be replaced at all points where this value is encountered in the original image, by the certain value $I_l$). The reliable identification of structuring requires [11(a,b)], however, a variable (i.e., «breathing», «dynamic») map -- to avoid artifacts and select an optimal final image. The best would be having an optimal map for each patch of the image (we call this a mosaic MDC). This, however, makes the image either unreasonably complicated, because of boundary effects, or requesting additional, non-contrasting procedures. Therefore we present here the images obtained with a homogeneous map of contrasting and additionally, in the special windows (see Figs. 2,8), we give a schematic drawing of the structuring that has been revealed through the mosaic MDC processing, to help a reader in recognizing the structuring in the presented image.

The use of MDC method enabled [6] to (i) observe the long-lived skeletal structures (SSs) in plasma, (ii) reveal the continuity of SS in the core and periphery, (iii) roughly trace dynamics of SSs during the entire discharge, (iv) resolve the fine structure of SS. The properties of skeletal structures are found to be much different from chaotic filaments. The observed longevity of SSs means that they may exist during the time period which largely exceeds the lifetimes predicted by the MHD approach and plasma kinetics (for the slow Z-pinch E-2 the lifetime of the above-mentioned straight blocks exceeded respective predictions by the orders of magnitude). The phenomenon of longevity for the case of the fast Z-pinch is illustrated with Figure 1 which shows two successive images (positives) with interframe time ~ 15 ns, taken in soft x-rays in a single discharge by two different electronic optical converters from almost the same position (the observation angles differ by the few degrees only). Both images are analyzed with respect to probable existence of objects which conserve their structure during the interframe period (note that the relative position of such objects may differ due to internal restructuring or global rotation of the entire structure). Such structures are found in the windows 1-4 where the original image is processed with the MDC method. Also, typical recognizable dark spots are marked (see labels 5 and 6). All the windows 1-4 (and respectively 1'-4') exhibit the presence of the same, conservative and quite distinctive structuring that indicates on the longevity of these structures.

The fine structure of skeletal structuring, whose recognizability allows tracing the dynamics of discharge, is illustrated with Fig. 2 which shows magnified images of windows 2 and 3 from Fig. 1. The windows show tubular structures of various diameter and declination with respect to observer. The tubules in the window 2 seems to have a thick bright region in the base of, presumably, a larger coaxial structure. The tubule in the window 3 has a small bright point on the front edge. For each image, we give schematic drawing of the structuring, which is obtained with the help of mosaic MDC method.

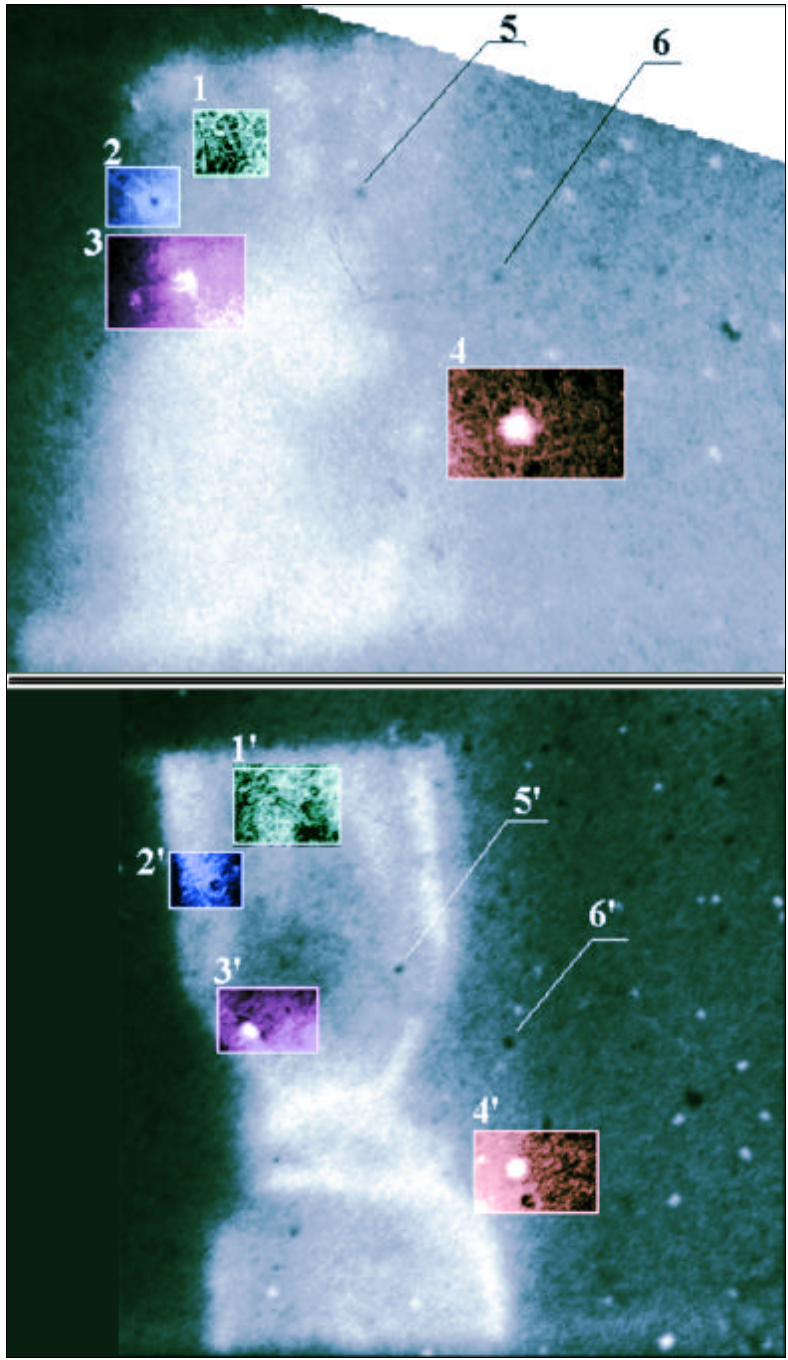

Fig. 1. Two successive images (positives, exposure 3 ns) with interframe time ~ 15 ns, taken in the Z-pinch facility S-300 in soft x-rays in a single discharge by two different electronic optical converters from almost the same position. Figure height ~ 1 cm. Detailed comparison of the images suggests the longevity of skeletal structuring.

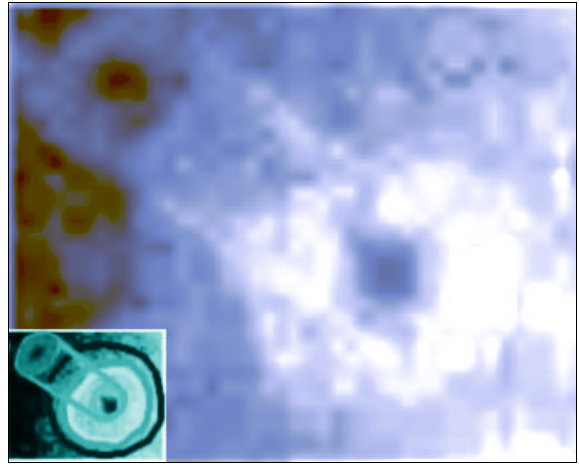

(a)

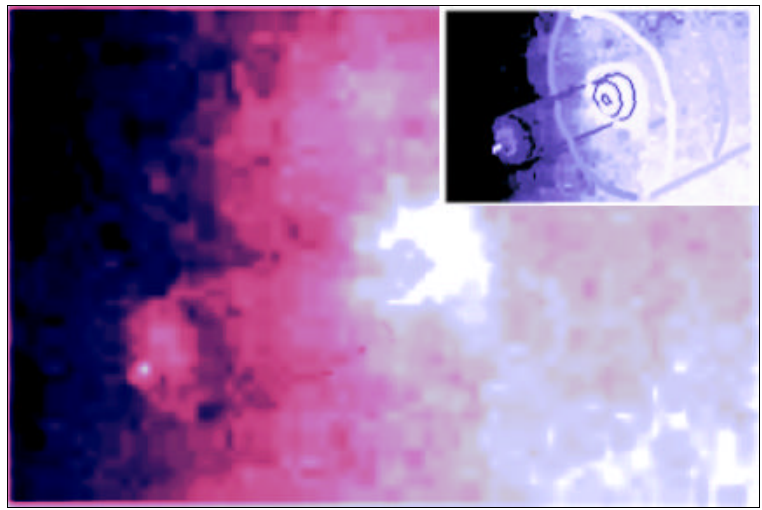

(b)

Fig. 2. The magnified images of the windows 2 and 3 from Fig. 1 (figures (a) and (b), respectively). The width of the images is ~1 mm (a) and 1.75 mm (b). Schematic drawings of the structuring, given in the corner windows, are obtained with the help of mosaic MDC method (see Sec. 2.1).

### 2.2 Skeletal Structuring in Dust Deposits in Tokamak

One of the most direct verification of hypotheses [1,2] for the presence of a self-similar skeletons made of a condensed matter may come from an analysis of dust deposits in electric discharges. This seems to be the case for the studies of various types of dust deposit in tokamak T-10. The presence of skeletal structures in the range 10 nm - 10 µm was found [5] in the images obtained by the high-resolution transmission (TEM) and scanning (SEM) electron microscopy of dust particles and thin films. The original database has been produced within the frame of the safety and environmental program of the ITER project (see the survey [12]). These studies were aimed, in particular, at analyzing the origin of unusual, non-trivial structures in the dust deposits (e.g., those of a cauliflower-like form [13] observed in tokamaks TEXTOR [14] and T-10 [12]). The enhancement of spatial resolution up to ~10 nm, via using the transmission electron microscopes, allowed [12] to find, besides cauliflower-like structures, new types of structuring originated from the presence of an internal dendritic skeleton, composed of nanotubular blocks, in certain types of dust particles and films.

In the literature, the aggregation of cauliflower-like structures is interpreted in different ways. The concept of dusty plasmas suggests (see e.g. [15]) that in fusion devices, similarly to plasma processing devices, such an aggregation takes place in the peripheral plasma and results from the action of the ambient plasma on the highly-charged dust microparticles. Alternatively, experiments on modeling the plasma-surface interaction in fusion devices, via studying the interaction of an ion beam with a solid target, suggest such a structuring to take place essentially on the surface of fusion facility's wall [12,16].

Here, we make a stress on the progress in appending the former evidences [5,3] with those from [7(a,b)] to show that the skeletal structures in the dust deposits may posses a dendritic form and that such skeletons may be hidden in an amorphous component (in tokamak T-10, this component is formed mostly by the hydrocarbons). Such a component may hide the internal skeleton either fully (to give a solitary dust particle, e.g. of submicron size) or partly (to give an agglomerate of visually separate particles (AVSP)). The example of AVSP may be found in Figures 4-6 in [7(a)]. It follows from analysis of AVSPs that the skeletal blocks inside AVSP may be arranged in a more or less ordered unified skeleton. The visually-separate quasi-spherical formations appear to be the blobs of an amorphous component trapped by (or deposited on) basic blocks of a dendritic skeleton. Roughly speaking, the quantity of amorphous component in this dust particle seems to be not sufficient to fill in the gaps between the neighboring blocks of the skeleton.

Note that compatibility of dendricity and tubularity in the same tubular skeleton [7(a,b)] supports the hypothesis [1,2] for a streamer-like mechanism of self-assembling of macroscopic tubular skeletons from nanotubular blocks during electric breakdown.

**2.3 Skeletal Structuring in Laser-Produced Plasmas**

Tracing the dynamics of self-assembling of skeletal structures in laboratory electric discharges would require high time resolution of imaging the initial stage of discharge to check not only the assembling as itself, but also the conservation of topology in a growing/expanding skeleton. These is also an implicit way to study the skeletal structuring, namely an analysis of time-integrated images of a plasma system where the formation of skeletal structures is expected to happen. In the latter case, if the growth/expansion of a skeletal structure during observation time is slow enough or the expansion slows down, i.e. saturates, the structure might be seen on the background of the time-integrated image of the expanding plasma component of the discharge [4].

The skeletal structures are expected to be formed even in such a bursty system like the plasma corona produced by the irradiation of a solid foils by a pulsed laser beam. Indeed, the expanding plasma may produce the electric current directed opposite to the laser pulse. This pulsed current produces a strong magnetic field which, in turn, may produce a return electric current and thus close an effective electric circuit at the periphery of the plasma corona. Regardless of specific kinetic mechanism of the above qualitative picture, the central near-axis region of the expanding plasma corona may possess the properties of a Z-pinch plasma column. Therefore, one could expect to observe some properties of a Z-pinch column in certain regions of the laser-produced plasma corona [4]. Indeed, the expected examples of transverse-to-axis straight filaments have been found in respective experimental data (see [3]).

Here, we present an exciting example of unusual fine structuring -- a sort of the walking-stick with toroid-like handles (see e.g. the end knob) -- seen as a three dimensional formation, thanks to the presence of above-mentioned handles and geometric perspective of a straight formation declined with respect to image's plane (Fig. 4). Moreover, it looks like there is a cartwheel-like structure located at this stick as a wheel at an axle.

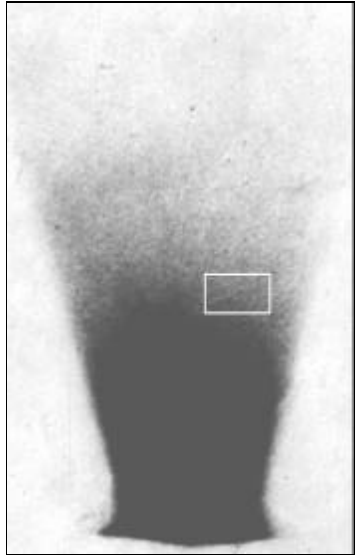

Fig. 3. The filtered X-ray pinhole image (negative) of the typical plasma corona produced by the Nd laser pulse ($\lambda$ = 1.06 μm, 2.5 ns duration, power density 5 $10^{13}$ W/cm$^2$) incident on the lavsan foil with aluminum dots. The original is taken from the database of experiments [17]. The target was located approximately on the bottom of the image, the beam was directed downward. Image height ~ 5 mm. The magnified image in the frame is shown in Figure 4.

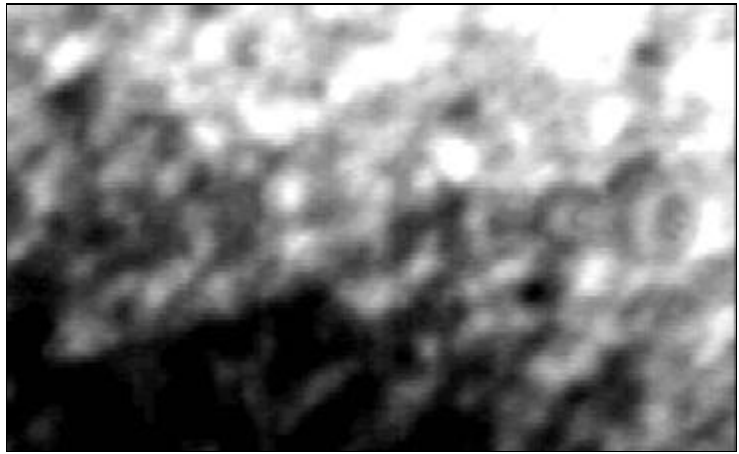

Fig. 4. A fragment of the image of Figure 3 marked with a frame. Image's width ~750 μm. One may see the image of a walking-stick whose end knob (of diameter ~ 80 μm) is seen on the right-hand side on the figure. This stick is oriented horizontally and declined with respect to figure's plane. One more handle-like structure of smaller diameter is seen on the stick behind the end knob. There is one more structure, though less distinct one, which is located on the stick, behind the latter structure - it looks like a cartwheel-like structure for which the stick serves as an axle-tree.

Another type of plasma dynamics which exhibits similar type of a skeletal structure, is found in the data from experiments [18] on laser irradiation of a fractal target, namely «agar-agar» ($C_{14}H_{18}O_7$). Here, in contrast to the case of an expanding plasma corona, major plasma is formed inside the fractal target. The transversely directed filamentary formation appears to possess a coaxial structure, namely a ring-shaped structures on a common thin axle (see the right-hand side of the bottom image in Fig. 5).

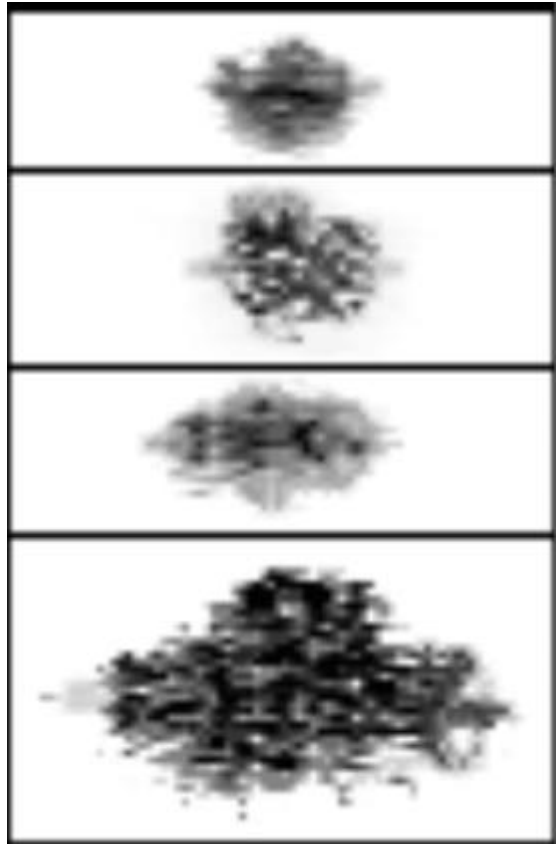

Fig. 5. The X-ray pinhole images (negatives) of the brightest part of a plasma produced by the Nd laser pulse ($l$ = 1.054 μm, 2.5 ns duration, power density 5 $10^{13}$ W/cm$^2$) in the inner part of the «agar-agar» ($C_{14}H_{18}O_7$) film target of thickness 0.5 mm and mass density 2 mg/cm$^3$. The originals are taken from the database of experiments [18]. The upper edge of the target was located approximately on the top of each image, the beam was directed downward. The images are taken through aluminum filters. Four time-integrated images correspond to different thickness of these filters, in the range 5 μm to 20 μm, with the thinnest one for the bottom image. Image's width is 720 μm. Spatial resolution is 20 μm. A straight filament directed orthogonal to laser pulse direction is seen on all the images whereas the elliptical images of the circular filaments centered around the above-mentioned filament are distinct mostly for the thinnest filter (see right-hand side of the bottom image).

For more examples of skeletal structures in laser-produced plasmas see ref. [4].

The original data from experiments [17,18] (Figures 3,5) have been kindly presented by N.G. Kovalskyi and V.V. Gavrilov.

## 3. ON DIAGNOSING FILAMENTARY STRUCTURES IN FUSION PLASMAS

The role of high time resolution for the identification of skeletal structures (and tracing their dynamics) was discussed in Sec. 2.1 for the case of the fast Z-pinch. The problem of diagnosing the fine structure of filaments in tokamaks has been discussed in [3] with regard to severe limitations imposed by the fast toroidal/poloidal rotation of both the plasma and skeletal components. It appeared that time resolution (~15 µs) sufficient for identification of skeletal structures at the very initial stage of discharge in tokamak T-6 (in particular, at a time t ~300 µs *before* appearance of the plasma current signal in Rogowski coil), via visible light imaging with the help of an electronic optical converter, has to be enhanced by an order of magnitude to identify similar structures at the steady-state stage of discharge, because of the fast rotation of plasma at this stage. Even the best available diagnostics of filaments in tokamak plasmas, namely ultra-high speed CCD camera used at tokamaks NSTX and Alcator C-Mod (see [19] and references therein), has space-time resolution still insufficient to detect fine structure of the filaments observed. Such a diagnostics allows tracing the dynamics of the few centimeters structuring [19] whereas the fine structure of filaments to be detected in the periphery of tokamaks TM-2, T-4, T-6 and T-10 was shown (see [3] and references therein) to require at least a few millimeters resolution.

The uniqueness of skeletal blocks, particularly the distinct layout of bright spots («hot spots») within a skeletal structure, makes it possible to measure the dynamic characteristics of both the skeletal and plasma components in various fusion devices. Here, the best way would be the high time resolution diagnostic with short enough interframe time (e.g. via imaging with the help of an electronic optical converter). There is also a simple way to use streak camera imaging whose opportunities have been shown [8] to be missed. This follows from an analysis of the data on streak camera imaging, with effective time resolution of the order of one microsecond, in the visible light in the former experiments in tokamaks TM-2, T-4, and T-6. The streak camera imaging in tokamaks provides diagnostic opportunities based on the existence of a long-lived fine structure of filaments (of luminosity) in plasmas (here the longevity implies that the structure is identifiable during the time period comparable with the slow MHD times, e.g., with the period of plasma toroidal rotation). The conclusions of analysis [8] are as follows.

First, skeletal structures do rotate in tokamaks, probably at a speed close to that of plasma. Indeed, just the relative motion, with respect to the diagnostic slit, of both the plasma column and the internal skeletons makes it possible to obtain a side-on image of the plasma column (of course, under condition of simultaneous motion, with respect to the slit, of the exposed film, or under similar conditions for an electronic reading of the signal). Thanks to plasma's radiation emission, the image on the film is produced in regime of a self-scanning of electric discharge's interior. Note that rotation of the plasma is visible because of perturbations of radiation emissivity of plasma in the regions of density/temperature perturbations (at the plasma column periphery, these perturbations are seen in the visible light and practically coincide with magnetic perturbations measured with magnetic probes). The fine structure of streak camera's images taken in tokamak T-6 (major/minor radii 70/20 cm; $B_T \sim 0.9$ T; $I_p \sim 100$ kA; $T_e(0) \sim 0.4$ keV, $n_e(0) \sim 2 \; 10^{13} \, cm^{-3}$) is illustrated with Figures 6-8.

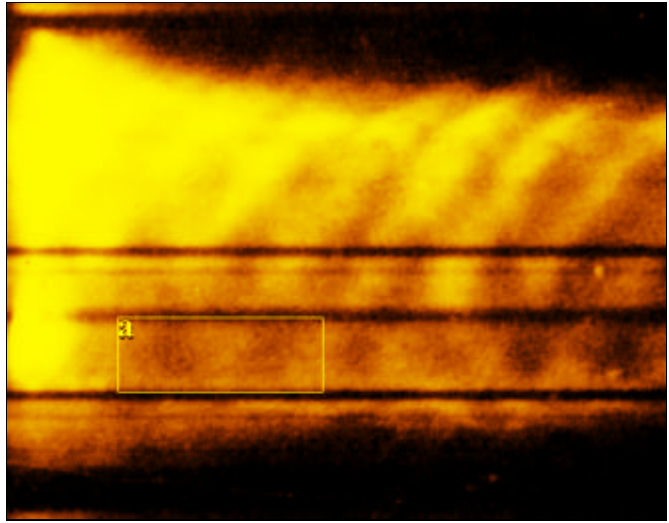

Fig. 6. A fragment of the streak camera image (positive, height 40 cm)) of the plasma column (toroidal direction is horizontal) taken in tokamak T-6 through a vertical slit. Thick horizontal lines are the reference marks (namely, the wires) on the outer surface of diagnostic window.

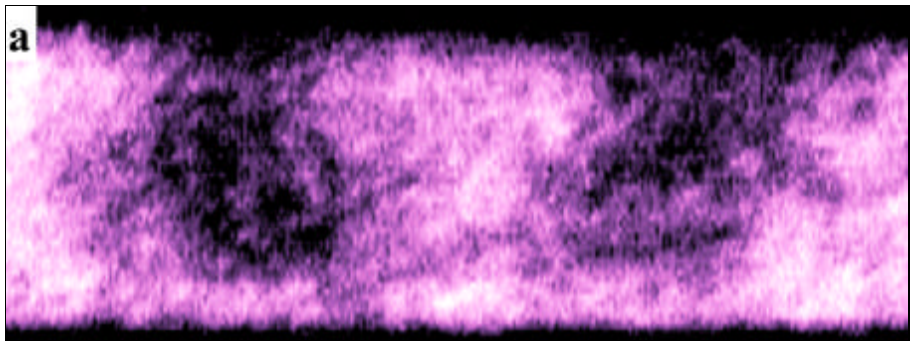

Fig. 7. The fragment of the image of Fig. 6 marked with a window. The original of Fig. 6 is only slightly processed here with the method of multilevel dynamical contrasting (MDC) [11(a,b)]. The width of the image is evaluated to be ~15 cm (the velocity of the exposed film motion was empirically adjusted to minimize the blurring of bright spots in toroidal direction in the given type of tokamak discharge; this allowed to evaluate the horizontal length scale in the image, for the given optical scheme of imaging). The perspective in the images of straight filaments may be seen as a continuous, smooth screening of the filament by the ambient plasma.

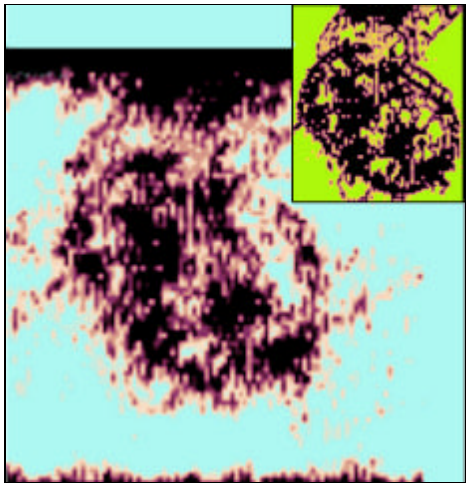

Fig. 8. The left part of the image in Fig. 7 which is processed with the MDC method. The scheme in the right corner insert is obtained with the help of mosaic MDC method (see Sec. 2.1). Thanks to empirically adjusted coincidence of toroidal rotation of the skeletal structures and (respectively scaled, for the given optical scheme) velocity of the exposed film, the blurring of bright spots takes place only in poloidal direction. In such a situation, both torodial and poloidal rotation velocities of bright spots (and skeletal structures) have been evaluated [8].

Second, a new method of streak camera imaging was suggested [8] which allows the determination of time evolution of direction and magnitude of rotation velocity of skeletal structures in a tokamak. This exploits the «blurring» of bright spots (within these filaments) in toroidal and poloidal directions, for a wide enough slit. The above determination in a *single* discharge requires a special optical scheme and preliminary optimization, trying only *few* discharges, of the time-dependent velocity of pulling the exposed film (or its electronic analog). The examples are presented [8] of evaluating the velocities from the data from tokamak T-6 experiments in the ordinary scheme which requires trying *many* discharges to optimize the imaging.

## 4. SKELETAL STRUCTURES AT GEO- AND ASTROPHYSICAL LENGTH SCALES

We try to draw a bridge between laboratory experiments and space with presenting a short gallery of cartwheel-like structures -- probably the most inconvenient objects for being described universally for the *entire* range of length scales under consideration [9]. The cartwheels of dramatically different size have been found (see [9]) in the following examples of (i) big icy particles of a hail, of several centimeters in diameter, (ii) a fragment of tornado of estimated diameter of some hundred meters, (iii) a fragment of solar coronal mass ejection of estimated diameter ~5 $10^{11}$ cm, and (iv) supernova remnant, «a flaming cosmic wheel» forty light-years across, i.e. ~4 $10^{19}$ cm.

We may add to the last item of this list the wheel-like supernova remnant G11.2-0.3 which is also 40 light years across (see [23] .../cycle1/1227/1227_xray.tif), and a two-ring coaxial structure, with the inner ring of one light year in diameter, in the centre of the Crab nebula (see [23] .../0052/0052_x-ray_lg.jpg, where radial structures are less distinct though). The gallery of cosmic wheels is obviously to be finished with the object named by the astronomers the Cartwheel galaxy, 150,000 light years across, i.e. ~1.4 $10^{23}$ cm (see [24] .../gif/cartknot.gif).

Significantly, most distinct example of cosmic wheel's skeleton (Fig. 4 in [9]) tends to repeat the structure of the cartwheel, coated with ice, (Fig. 1 in [9]) up to details of its constituent blocks. For instance, some of radially directed spokes are ended with a tubular structure seen on the outer edge of the cartwheel. Moreover, there are evidences for the trend toward self-similarity because, e.g., the cartwheel structure of the icy particle contains a smaller cartwheel as a constituent block (see the edge cross-section of the tubule in the left lower window in Fig. 1 in [9]). This fact extends to larger length scales the evidences for such a trend in tubular skeletons in the dust deposits [5,7].

The ring-shaped structure of supernova remnant may be reproduced within the frame of hydrodynamic description of an expanding magnetized plasma-gaseous medium (see, e.g., [25] for the formation of the three-ring structure around supernova 1987A). However, we didn't find in the literature an attempt to model a ring-shaped system with radial bonds, i.e. the cartwheel.

Besides the distinctive topology (e.g., the cartwheels) of general layout of bright spots within skeletal structures, another evidence for the phenomenon of skeletons comes from the resolution of fine structure of luminosity around, at first glance, solitary bright spots. The best relevant evidence seems to be the shining edge of a truncated straight filament which belongs to a skeletal network (see [9]). The similarity of an electric torch-like structure of bright spots at different length scales is found in various laboratory electric discharges and in space, in the range $10^{-2}$-$10^{22}$ cm. Such a similarity suggests that the blocks of skeletons may work as a guiding system for (and/or a conductor of) electromagnetic signals. Therefore, the open end of a dendritic electric circuit or a local disruption of such a circuit (e.g., its sparkling, fractures, etc.) may self-illuminate it to make it observable.

Some indications on the probable presence of skeletal structuring at cosmological lengths, namely in the clusters of galaxies in the range $10^{24}$-$10^{26}$ cm, are found [26] in analysing the galaxy redshift surveys.

## 5. THE PRESENT STATUS OF HYPOTHESES [1,2]

The prediction [1,2] of the phenomenon of skeletons in a wide range of lengths, starting from nanoscale structures, was based on appealing to exceptional electrodynamic properties of their hypothetical building blocks -- first of all, the ability of these blocks to facilitate the electric breakdown in laboratory discharges and to assemble macroskeletons. The carbon nanotube (CNT) or similar nanostructures of other chemical elements have been suggested to be such blocks. The self-assembling of skeletons was suggested to be based dominantly on *magnetic* phenomena. This, in fact, assumes the following simultaneous processes, namely

($\alpha$) externally driven expansion/inflation of a self-assembling network (in laboratory discharges, there is an inflow of magnetic field from the external electric circuit, which is especially intense at initial stage of discharge; in space, an expansion may result, e.g., from nuclear energy release produced by gravitational instability and collapse),

($\beta$) sticking of the blocks together -- due to trapping of magnetic flux by CNTs, and/or their assemblies, and respective mutual magnetic dipole attraction; at macroscopic scales this mechanism agrees with the experimentally verified trend [27] in magnetically confined plasmas to form the so-called force-free configuration which sustains a balance between longitudinal and transverse magnetic confinement, and thus always produces a longitudinal attraction both in the entire plasma column and individual electric current filament, and

($\gamma$) *partial* solidification due to welding of blocks by the passing electric current.

We believe that the mechanisms ($\alpha$),($\beta$), and ($\gamma$) may give a parachute-like expansion of a dendritic network (namely, a parachute with the «liquid» strops and the localised explosive sources of dendricity), which selects the structures of a matter-saving and survivable geometry (first of all, tubules and cartwheels, and their combinations) regardless of specific source of expansion/inflation.

The indications on plausibility of the anomalous magnetism and, in particular, on the ability of CNTs, and/or their assemblies, to trap and almost dissipationlessly hold the magnetic flux, with the specific magnetization high enough to stick the CNTs together, come from observations of superconductor-like diamagnetism in the assemblies of CNTs at high enough temperatures. Such evidences are obtained for the self-assemblies of CNTs (which contain, in particular, the ring-shaped structures of few tens of microns in diameter) inside *non-processed* fragments of cathode deposits, at room temperatures, [28] and for the artificial assemblies, at 400 K [29].

The indications on the conservation of topology in a growing/expanding skeletal structure (i.e. on the compatibility of rigidity of certain blocks with the global/local expansion) come from the available data -- at this point, unfortunately, rare ones -- on tracing the *dynamics* of a cosmic explosion, namely such in the young star system XZ Tauri, ~0.01 light years across (.../PR/2000/32/ [24]). Here, we found the signs of a two-dimensional expansion (it's worth to call this a «planar explosion»). This example seems to be an inflationary production of the cartwheel-like structures on the common axle and is compatible with the interplay of the mechanisms (α) and (β).Another example (though, a static one) of a distinct planar structuring is the «large thin equatorial disk» (.../PRC96-23a [24]) less than one light year across around the star Eta Carinae (.../EtaCarC.jpg [24]) which has been released as a supernova explosion.

At this point, the signs of the saturation of the growth/expansion of a dendritic network are only implicit ones. For instance, the revealed phenomenon [9] of an open, «shining» ends of the network could result not only from local disruption of the network but also from local termination of a dendritic growth because of local exhaust of the proper material. Another argument in favour of saturation is the presence of typical skeletal structures in the *time-integrated* x-ray images of one of the most bursty phenomenon in the laboratory, namely plasma corona produced by the irradiation of a solid target with a short laser pulse (see Sec. 2.3).

Regarding all the above-mentioned evidences for the dendricity of skeletal structures, we have to note that the typical block of skeletons, namely the cartwheel on an axle and the tubule with the central rod and the cartwheel in the edge cross-section, are topologically the dendrites. An example of simultaneous dendricity and tubularity of the skeleton composed of tubular nanofibers is found in the submicron dust particle (see Sec. 2.2 and [7(a,b)]).

The solution to the problem of survivability of skeleton in hot plasmas has been suggested within the framework of the problem of non-local (non-diffusive) transport of heat observed in high-temperature plasmas for controlled thermonuclear fusion (cf. the survey [30]). The microsolid skeletons were suggested [31] to be self-protected from an ambient high-temperature plasma by thin vacuum channels self-consistently sustained around the skeletons by the pressure of high-frequency (HF) electromagnetic waves, thanks to the skeleton-induced conversion of a small part of the incoming slow, quasi-static magnetic field (poloidal, in tokamaks, or azimuthal, in Z-pinches) into HF waves of the TEM type (a «wild cable» model, see P2_028 in Ref. [31]). This model allows to evaluate the width and length of vacuum channels around straight blocks of skeletons from the measurements of HF electric fields, both inside and outside the plasma column. The respective results [31,32] for the case of tokamak T-10 and gaseous Z-pinch reasonably agree with visible dimensions of observed straight blocks of skeletons.

## 6. CONCLUSIONS. PROBABLE ROLE OF A FRACTAL DUST IN FUSION DEVICES. SOME IMPLICATIONS OUTSIDE FUSION SCIENCE.

We start with a conclusion related to basic physics. The following two observational facts revealed in the range $10^{-5}$-$10^{23}$ cm, namely

(i) topological identity (i.e. the similarity) of skeletal structures (especially of the «cartwheel» as a structure of essentially non-hydrodynamic nature), and

(ii) a trend of assembling bigger structures from similar smaller ones (i.e. the self-similarity),

suggest all these skeletal structures, similarly to skeletons in the particles of dust and hail, to possess a fractal condensed matter of particular topology of the fractal [9]. Specifically, this matter may be self-assembled from nanotubular blocks in a way similar to that in the skeletons found in the submicron dust particles [5,7].

The probable implications of this hypothesis <u>inside the fusion science</u> are based on the probable positive role of skeletal structures. These include, in particular, the following problems related to, respectively, creation and maintenance of relevant regimes of electric discharge in fusion devices.

(1) The practical necessity to attain higher reproducibility/predictability of the very initial stage of discharge in fusion devices . These include:

• facilitation of electric breakdown of the working gas in the discharge chamber (e.g., extension of the stimulating role of graphite and graphite composites in electric breakdown phenomena, as identified in various applications, to the case of conditions peculiar to fusion devices),

• identification of microscopic mechanisms responsible for the existing algorithms -- at best, semi-empirical ones -- of attaining the successful regimes of electric discharge in fusion devices.

(2) The second problem assumes improving the plasma thermal energy confinement. Here, an application of the «wild cable» model [31] to particular diagnostic algorithms of plasmas in fusion devices may contribute to resolving the following significant problems :

• probable control of the nonlocal, non-diffusional component of heat transport in tokamaks, which has been observed in experiments on the fine resolution of heat pulse propagation from an instant point source of the heat or cold (cf., e.g., the «cold pulse» experiments);

• further optimization of experiment schemes in the wire-array fast Z-pinches and identification of physical roots of the still unresolved discrepancy between the visible dynamics of plasma implosion and the unexpectedly high values of parameters of the soft x-ray radiation pulse (namely, unpredictably high radiation yield and especially high radiation power, unpredictably high efficiency of conversion of discharge's magnetic energy into radiation pulse, etc.).

The probable implications of this hypothesis <u>outside the fusion science</u> are suggested by the manifestations of the phenomenon of skeletal structuring at various length scales. These include:

(1) Material science.

Skeletal structures observed in high-current discharges in laboratory may be the prototype of a new nanomaterial which might be of interest for a wide range of potential industrial applications. (Recall that nanomaterial is such whose major properties are determined by its constituent blocks whose size lies in the nanometer range).

(2) Geophysics.

Identification of skeletal structuring in severe weather phenomena (tornado, etc.) may improve the state of the science in this field. The importance of such an analysis is suggested by the fact that the mechanism which triggers the tornado is not identified yet. Specifically, it is not possible to differentiate between tornadic and non-tornadic mesocyclones and, correspondingly, predict the tornado event up to the very birth of this monster.

(3) Astrophysics and cosmology (see [26]).

(3a) Certain mechanical strength (rigidity) of skeletal structures at galactic and extragalactic length scales may eliminate the necessity to introduce a «dark matter».

(3b) Purely gravitational description of large-scale structure of the Universe is likely to be appended with a contribution of hypothetical skeletons composed from nanotubular structures (note that the integrity of such skeletons is governed by the quantum electrodynamics; in particular, chemical bonds are responsible for the integrity of carbon nantotubes).

(3c) A combination of two observational facts – namely, the presence of skeletal structures in the range $10^{-5}$ cm - $10^{23}$ cm [9] and the signs of skeletal structuring in the range $10^{24}$-$10^{26}$ cm [26], as suggested by an analysis of the redshift surveys of galaxies and quasars – hints at the presence of a baryonic cold skeleton of the Universe. This hypothesis may be shown to have no conflict with the ultrahigh isotropy of cosmic microwave background radiation and the high enough uniformity of Hubble's expansion of the Universe.

**Acknowledgments**

We thank all our colleagues who presented their original databases and collaborated with us at various stages of the present research, especially to B.N. Kolbasov and P.V. Romanov. Our special thanks to V.I. Kogan for his support.

Partial financial support from the Russian Federation Ministry for Atomic Energy and the Russian Foundation for Basic Research is acknowledged.